\documentclass[aps,preprint,prd,nofootinbib]{revtex4-1}
\usepackage[utf8]{inputenc}
\usepackage{booktabs}
\usepackage{slashed}
\usepackage{indentfirst}
\usepackage{graphicx}
\usepackage{amsfonts,amssymb,amsmath}
\usepackage{graphicx}
\usepackage{psfrag}
\usepackage{psfrag}
\usepackage{multirow}       
\usepackage{url}
\usepackage[hang]{subfigure}
\usepackage{supertabular}
\usepackage{longtable}
\usepackage[colorlinks=false]{hyperref}
\usepackage[amsmath,thmmarks,hyperref]{ntheorem}
\usepackage{soul}

\begin{document}

\title{Strong decays of $D_J(3000)$ and $D_{sJ}(3040)$}
\author{Si-Chen Li$^{[1]}$, Tianhong Wang$^{[1]}$\footnote{thwang@hit.edu.cn}, Yue Jiang$^{[1]}$, Xiao-Ze Tan$^{[1]}$, \\Qiang Li$^{[1]}$, Guo-Li Wang$^{[1]}$, Chao-Hsi Chang$^{[2,3]}$}
\address{$^1$Department of Physics, Harbin Institute of Technology, Harbin, 150001 \\
$^2$CCAST(World Laboratory), P.O. Box 8730, Beijing 100080, People's Republic of China\\
$^3$Institute of Theoretical Physics, Chinese Academy of Sciences, P.O. Box 2735, Beijing 100080, People's Republic of China
}

 \baselineskip=20pt

\begin{abstract}
In this paper, we systematically calculate two-body strong decays of newly observed $D_J(3000)$ and $D_{sJ}(3040)$ with 2P$(1^+)$ and 2P$(1^{+\prime})$ assignments in an instantaneous approximation of the Bethe-Salpeter equation method. Our results show that both resonances can be explained as the 2P$(1^{+'})$ with broad width via $^3P_1$ and $^1P_1$ mixing in $D$ and $D_s$ families. For $D_J(3000)$, the total width is 229.6 MeV in our calculation, close to the upper limit of experimental data, and the dominant decay channels are $D_2^*\pi$, $D^*\pi$, and $D^*(2600)\pi$. For $D_{sJ}(3040)$, the total width is 157.4 MeV in our calculation, close to the lower limit of experimental data, and the dominant channels are $D^*K$ and $D^*K^*$. These results are consistent with observed channels in experiments. Given the very little information that has been obtained from experiments and the large error bars of the total decay widths, we recommend the detection of dominant channels in our calculation.

 \vspace*{0.5cm}


\end{abstract}

\maketitle

\section{INTRODUCTION}
Recently, great progress has been made in $D$ and $D_s$ families~\cite{Chen16}. Numerous highly excited states have been found in experiments. These states stimulate great interest and provide a good platform to study nonperturbative QCD. In the spectrum of the 2P wave, we notice that no 2P states have been confirmed in experiments yet in charmed and charm-strange families. The study of these newly discovered resonances can enlarge our knowledge of spectroscopy and also the properties of 2P states.    

In the charm-strange family, $D_{s1}^*(2700)$ was discovered by Belle in 2008 with a $1^-$ quantum number \cite{Ds2700}; $D_{s1}^*(2860)$ and $D_{s3}^*(2860)$ were observed by LHCb in 2014 with $1^-$ and $3^-$ quantum numbers, respectively \cite{Ds2860}. In 2009, $D_{sJ}(3040)$ was reported by the BABAR Collaboration in the $D^*K$ channel \cite{Ds3040}. In the charmed family, BABAR announced four resonances in 2010, namely, $D(2550)$, $D^*(2600)$, $D(2750)$, and $D^*(2760)$ \cite{D2550etc}. By analyzing the helicity distribution, the first two are identified as a 2S doublet with unnatural and natural parity, while the latter two are good candidates for D-wave states; the assumption corresponds to their strong decays in theoretical calculations \cite{D2550identification}. In 2013, the LHCb Collaboration announced two resonances, $D_J(3000)$ and $D_J^*(3000)$, with unnatural and natural parities, respectively, through the $D^*\pi$ and $D\pi$ channels \cite{D3000}. In 2016, LHCb announced two new resonances~\cite{D3214}, namely, $D_3^*(2760)$ and $D_2^*(3000)$, which have $3^-$ and $2^+$ quantum numbers.  

In our previous work \cite{Wangstrongdecay}, the strong decays of $3^-$ states like $D_{s3}^*(2860)$ and $D_3^*(2760)$ have been analyzed. Some $1^-$ states like $D_{s1}^*(2700)$, $D_{s1}^*(2860)$, $D^*(2600)$, $D^*(2650)$, $D_1^*(2680)$, and $D_1^*(2760)$ have been investigated through strong decays \cite{Liqiangstrongdecay}. The $1^-$ state is a mixture of $^3S_1$ and $^3D_1$ waves. By fitting the experimental branching ratios, the mixing angles between 2$^3S_1$ and 1$^3D_1$ states for charmed and charm-strange families are discussed. Among these new resonances, two resonances we have not discussed yet are $D_{sJ}(3040)$ and $D_J(3000)$. They are good candidates for the 2P$(1^{+})$ states and are measured as \cite{Ds3040,D3000} 
\begin{eqnarray}
\begin{aligned}
m_{D_{sJ}(3040)^{+}}=\left(3044\pm8^{+30}_{-5}\right)\ \mathrm{MeV},\\
\varGamma_{D_{sJ}(3040)^{+}} =\left(239\pm35^{+46}_{-42}\right)\ \mathrm{MeV},\\
m_{D_{J}(3000)^{0}}=\left(2971.8\pm8.7\right)\ \mathrm{MeV},\\
\varGamma_{D_{J}(3000)^{0}} =\left(188.1\pm44.8\right)\ \mathrm{MeV}.\\
\end{aligned}
\end{eqnarray}    
They have unnatural parity and thus are $0^-$, $1^+$, $2^-$, $3^+$, $\cdots$ states. Their masses are around 3000 MeV, lower than the $3^1S_0$ and higher than the $1^1D_2$ and $1^3D_2$ states in theoretical predictions, located in the mass region of 2P$(1^{+})$ states \cite{Godfreyresults}. Therefore, the assignments of the 2P$(1^+)$ states are reasonable. In addition, by studying the semileptonic decay of $B$ and $B_s$ mesons, these two candidates can also be interpreted as 2P$(1^+)$ states \cite{simi1,simi2,simi3}.

We notice that very little decay channels are given in experiments, and there should be many more decay channels. To identify their quantum numbers and determine their decay properties, we calculate the OZI-allowed two-body decay channels of the two new resonances with an instantaneous Bethe-Salpeter approach, which have been applied successfully in other strong decay channels and proved to be a good method \cite{bs1,bs2,bs3}. There should exist 2P$(1^+)$ and 2P$(1^{+'})$ states theoretically, while only one candidate has yet been observed in $D$ and $D_s$ families, respectively. The calculation can help us to search for the other state and to have a better understanding of the mixing angle between the $^1P_1$ and $^3P_1$ states as well.

We present a phenomenological analysis of the two candidates. We use a reduction formula, PCAC, and low energy theorem to deal with the case of a pseudoscalar final light meson. Since it is not valid for vector light meson such as $K^*$ or $\rho$, we adopt the effective Lagrangian method to calculate the channels of the vector light meson.    

Apart from an instantaneous Bethe-Salpeter approach, several other methods can describe the form factor and hadronic transition, such as a nonrelativistic quark model \cite{liqiang27}; heavy effective theory \cite{liqiang28}; effective Lagrangian approach based on heavy quark chiral symmetry \cite{liqiang30}; Eichten, Hill, and Quigg (EHQ) decay formula \cite{liqiang24}; quark pair creation (QPC) models \cite{liqiang31}; lattice QCD \cite{LatticeQCDWTH}; QCD sum rules\cite{QCDsumruleWTH}; Dyson-Schwinger-equation approach\cite{DSequationWTH}; and AdS-QCD method\cite{AdSQCD}.

The paper is arranged as follows. In Sec. II, we present the theoretical formalism of strong decays. If the final light meson is a pseudoscalar meson, the quark-meson coupling is introduced by two methods, if the final light meson is a vector state, an effective Lagrangian method is adopted. In Sec. III, we give Bethe-Salpeter wave functions and their mixing. In section IV, we present our results of OZI-allowed two-body strong decays of these two heavy-light mesons and compare our results with those from other models. Finally, we give a summary in Sec. V.


\section{THE FORMALISM OF STRONG DECAY}
\begin{figure}[!hbt]
\centering
\includegraphics[height = 4.5cm, width = 10 cm]{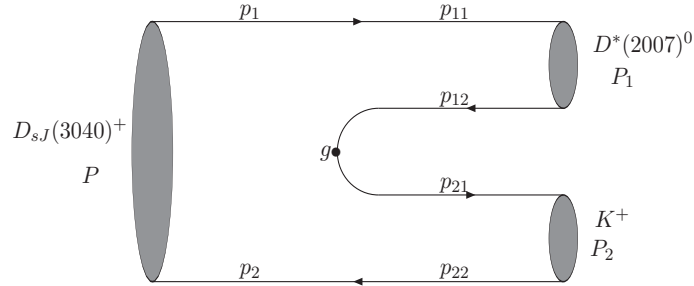}
\caption{$D_{sJ}(3040)^+$ decays to $D^*(2007)^0K^+$.}
\label{Feynmandiagram} 
\end{figure}

In this section, we show the process of calculating strong decays under the framework of an instantaneous Bethe-Salpeter equation. In order to illustrate how to apply our approach to strong decays, we take $D_{sJ}(3040)^+\rightarrow D^*(2007)^0K^+$ as an example. In the $^3P_0$ decay model, a quark-antiquark pair is created from the vacuum, the Feynman diagram of this process is given in Fig. \ref{Feynmandiagram}. 

The wave function of the final heavy meson can be obtained by solving corresponding instantaneous Bethe-Salpeter equation. By using the reduction formula, the transition matrix element of strong decay can be written as \cite{Liqiangstrongdecay}
\begin{eqnarray}
\begin{aligned}
&\left\langle D^*(2007)^0(P_1)K(P_2)|D_{sJ}(3040)^+(P)\right\rangle \\
=&\int {\rm d}^4x e^{iP_2\cdot x}(M_K^2-P_2^2) \left\langle D^*(2007)^0(P_1)|\Phi_K(x)|D_{sJ}(3040)^+(P)\right\rangle, 
\end{aligned}
\end{eqnarray}
where $P$ is the momentum of the initial meson, and $P_1$, $P_2$ are the momenta of the final heavy and light meson, respectively. $\Phi_K(x)$ is the light scalar meson field. By using the PCAC approximation method, the light scalar meson field can be expressed as \cite{Liqiangstrongdecay} 
\begin{eqnarray}
\begin{aligned}
\Phi_K(x)=\frac{1}{M_K^2f_K}\partial_{\xi}(\overline{u}\gamma^{\xi}\gamma^5s),
\end{aligned}
\end{eqnarray}
where $f_K$ is the decay constant of the $K$ meson. Inserting the above equation into Eq. (2), we get
\begin{eqnarray}
\begin{aligned}
&\left\langle D^*(2007)^0(P_1)K(P_2)|D_{sJ}(3040)^+(P)\right\rangle \\
=&\frac{M_K^2-P_2^2}{M_K^2f_K} \int {\rm d}^4x e^{iP_2\cdot x} \left\langle D^*(2007)^0(P_1)|\partial_{\xi}(\overline{u}\gamma^{\xi}\gamma^5s)|D_{sJ}(3040)^+(P)\right\rangle \\
=&\frac{-iP_{2\xi}(M_K^2-P_2^2)}{M_K^2f_K} \int {\rm d}^4x e^{iP_2\cdot x} \left\langle D^*(2007)^0(P_1)|\overline{u}\gamma^{\xi}\gamma^5s|D_{sJ}(3040)^+(P)\right\rangle.
\end{aligned}
\end{eqnarray}

Finally, by using the low energy theorem, the transition amplitude in the momentum space can be expressed as \cite{Liqiangstrongdecay}
\begin{eqnarray}
\begin{aligned}
\mathcal M \approx \frac{-iP_{2\xi}}{f_K}\left\langle D^*(2007)^0(P_1)|\overline{u}\gamma^{\xi}\gamma^5s|D_{sJ}(3040)^+(P)\right\rangle.
\end{aligned}
\end{eqnarray}

Apart from the approach with the reduction formula, PCAC approximation, and low energy theorem, we can also directly use the effective Lagrangian method to obtain the transition amplitude. The effective Lagrangian of this process is \cite{ZhongXianHui3040}
\begin{eqnarray}
\begin{aligned}
\mathcal {L}_{qqP}=\frac{g}{\sqrt{2}f_P}\overline{q}_i\gamma^{\xi}\gamma^5q_j \partial_{\xi}\phi_{ij},
\end{aligned}
\end{eqnarray}
where
\begin{gather*}
\phi_{ij}=\sqrt{2}\begin{bmatrix} \frac{1}{\sqrt{2}}\pi^0+\frac{1}{\sqrt{6}}\eta && \pi^+  && K^+ \\\pi^- && -\frac{1}{\sqrt{2}}\pi^0+\frac{1}{\sqrt{6}}\eta  && K^0 \\ K^- && K^0  && -\frac{2}{\sqrt{6}}\eta  \end{bmatrix} 
\end{gather*}
is the chiral field of the pseudoscalar meson. $g$ denotes the quark-meson coupling constant.

Within Mandelstam formalism, the transition amplitude can be expressed as the overlapping integral over the Salpeter wave functions of the initial and final mesons \cite{Wangstrongdecay}
\begin{eqnarray}
\begin{aligned}
\mathcal {M}&=\frac{-iP_{2\xi}}{f_K}\left\langle D^*(2007)^0(P_1)|\overline{u}\gamma^{\xi}\gamma^5s|D_{sJ}(3040)^+(P)\right\rangle\\
&\approx \frac{-iP_{2\xi}}{f_K} \int \frac{{\rm d}^3 \vec{q}}{(2\pi)^3} {\rm Tr} \left[  \overline{\varphi}^{++}_{P_1}(\vec{q}-\frac{m_1'}{m_1'+m_2'}\vec{P}_1 )\frac{\slashed{P}}{M}\varphi_P^{++}(\vec{q})\gamma^{\xi}\gamma^5  \right].
\end{aligned}
\label{eq:1} 
\end{eqnarray}

In the above formula, the only left unknown is the form of the Bethe-Salpeter wave functions of initial and final mesons, which will be given in detail in the next section.

If the final light meson of the strong decay is $\eta$ or $\eta'$, the mixing of the octet and singlet should be considered, and the mixing equation is 
\begin{gather}
\begin{bmatrix} \eta  \\ \eta' \end{bmatrix}=\begin{bmatrix} \cos{\theta_{\eta}} && \sin{\theta_{\eta}}  \\ -\sin{\theta_{\eta}} && \cos{\theta_{\eta}} \end{bmatrix}    \begin{bmatrix} \phi_8  \\ \phi_0 \end{bmatrix},
\end{gather}
where $\phi_8$ and $\phi_0$ are the flavor $SU(3)$ octet and singlet states, respectively. As in Ref.~\cite{Zhangsicheng}, we adopt the mixing angle $\theta_{\eta}\simeq19^{\circ}$. This value is achieved in Ref.~\cite{Choi} by using the light cone quark model. It is also a result of ChPT by considering higher order corrections~\cite{mixingetaoftreeandChPT} (The tree level result is $9.95^\circ$; see Refs.~\cite{mixingetaofmassdiagonalization,mixingetaoftreeandChPT}). Besides this, there is the masses of the mixing equation which links the mass of physical states and flavor states \cite{Wangstrongdecay}:
\begin{gather}
\begin{bmatrix} M^2_{\phi_8}  \\  M^2_{\phi_0} \end{bmatrix}=\begin{bmatrix} \cos^2{\theta_{\eta}} && \sin^2{\theta_{\eta}}  \\ \sin^2{\theta_{\eta}} && \cos^2{\theta_{\eta}} \end{bmatrix}    \begin{bmatrix} M^2_{\eta}  \\ M^2_{\eta'} \end{bmatrix}.
\end{gather}

An example involving $\eta$ is $D_{sJ}(3040)^+\rightarrow D_s^{*+} \eta$. Because the constitute quarks of $\phi_8$ and $\phi_0$ are $\frac{1}{\sqrt{6}}(d\overline{d}+u\overline{u}-2s\overline{s})$ and $\frac{1}{\sqrt{3}}(d\overline{d}+u\overline{u}+s\overline{s})$, the PCAC approximation relation in this decay is
\begin{eqnarray}
\begin{aligned}
\Phi_{\eta}&=\cos{\theta_{\eta}}\Phi_{\phi_8}(x)+\sin{\theta_{\eta}}\Phi_{\phi_0}(x)\\
           &=\frac{\cos{\theta_{\eta}}}{M^2_{\phi_8}f_{\phi_8}}\partial_{\xi}\left(\frac{\overline{u}\Gamma^{\xi} u+\overline{d}\Gamma^{\xi} d-2\overline{s}\Gamma^{\xi} s}{\sqrt{6}}\right)+\frac{\sin{\theta_{\eta}}}{M^2_{\phi_0}f_{\phi_0}}\partial_{\xi}\left(\frac{\overline{u}\Gamma^{\xi} u+\overline{d}\Gamma^{\xi} d+\overline{s}\Gamma^{\xi} s}{\sqrt{3}}\right)\\
           &=\left[\frac{-2\cos{\theta_{\eta}}}{\sqrt{6}M^2_{\phi_8}f_{\phi_8}}+\frac{\sin{\theta_{\eta}}}{\sqrt{3}M^2_{\phi_0}f_{\phi_0}}\right] \partial_{\xi}(\overline{s}\Gamma ^{\xi}s).\\
\end{aligned}
\end{eqnarray}
where $\Gamma^{\xi}$ is $\gamma^{\xi}\gamma^5$. Thus, the transition amplitude of this process can be written as 
\begin{eqnarray}
\begin{aligned}
\mathcal {M}=P_{2\xi}\left[  M_{\eta}^2  \frac{ -2\cos{\theta}}{\sqrt{6} M_{\phi_8}^2 f_{\phi_8}} +M_{\eta}^2    \frac{\sin{\theta}}{(\sqrt{3})M_{\phi_0}^2f_{\phi_0}}  \right]   \left\langle D_s^*|\overline{d}\gamma^{\xi}\gamma^5s|D_s(2{\rm P})\right\rangle.
\end{aligned}
\end{eqnarray}

In addition, there is also a mixing between $\pi^0$ and $\eta$ via $\phi_3$ and $\phi_8$, but because the mixing parameter is so small, we ignore the mixing between $\pi^0$ and $\eta$ \cite{phi38mixing}. Therefore, we treat $\pi^0$ and $\eta$ as pure states of $\phi_3$ and $\phi_8$, respectively.

If the final light meson is not a pseudoscalar but a vector meson, the PCAC cannot be applied. In this case, we use the effective Lagrangian method to get the transition amplitude. The Lagrangian of quark-meson coupling is \cite{ZhongXianHui3040}  
\begin{eqnarray}
\begin{aligned}
\mathcal L_{qqV}=\sum_j \overline{q}_j(a\gamma_{\mu}+\frac{ib}{2m_j}\sigma_{\mu\nu}P_2^{\nu})V^{\mu}q_j,
\end{aligned}
\end{eqnarray}
where $a=-3.0$ and $b=2.0$, representing the vector and tensor coupling strength, respectively; $\sigma_{\mu\nu}=\frac{i}{2}[\gamma_{\mu},\gamma_{\nu}]$; $V^{\mu}$ is the light meson field; and $m_j$ is the constitute quark mass of the final light meson. 
Therefore, the transition amplitude can be simplified as
\begin{eqnarray}
\begin{aligned}
\mathcal {M}=  \int \frac{{\rm d}^3 \vec{q}}{(2\pi)^3} {\rm Tr} \left[    \overline{\varphi}^{++}_{P_1}\frac{\slashed{P}}{M}\varphi_P\left(a\slashed{\epsilon}_2+\frac{ib}{4m_j}\left(\slashed{\epsilon}_2P_2-P_2\slashed{\epsilon}_2\right)\right)   \right].
\end{aligned}
\end{eqnarray}

Once we know the transition amplitude, the decay widths can be obtained by the following two-body decay formula
\begin{eqnarray}
\begin{aligned}
\Gamma=\frac{|\vec{P}_1|}{8\pi M^2}\frac{1}{2J+1} \sum\limits_{\lambda}|\mathcal M|^2,
\end{aligned}
\end{eqnarray}
where $P_1$ is the momentum of the final meson, $|\vec{P}_1|=\sqrt{[M^2-(M_1-M_2)^2][M^2-(M_1+M_2)^2]}/2M$, and $J$ is the quantum number of the total angular momentum of the initial meson. Under the assumption of the 2P$(1^+)$ states of these two new resonances, $J=1$.

\section{Bethe-Salpeter Wave Function}

In the last section, we show the processes for how we deal with different cases of strong decays and get the transition amplitude as well; the only thing left is the form of the Bethe-Salpeter wave function. In this section, we construct the Bethe-Salpeter wave function of different states for initial and final mesons and give the mixing equation of the $1^{+}$ states. It should be pointed out that compared with double heavy mesons, the use of instantaneous approximation of the Bethe-Salpeter equation for a heavy-light charmed meson is not very good. However, we still use this approximation here, as it makes the model have the same predictive power as other quark models on the one hand, and our previous work~\cite{FuCPC} with this model gets results that agree with experimental data on the other.

The instantaneous wave functions of mesons are constructed by the momenta, polarization vector (tensor), metric tensor, etc, which combine with gamma matrices to form covariant terms. For the states with quantum number $1^{+}$, there are eight independent covariant terms in general. Strictly speaking, one should solve the instantaneous Bethe-Salpeter equation to get the mass spectrum and corresponding wave functions of the $1^+$ and $1^{+\prime}$ states at the same time. But here in order to compare with other quark models, we solve the equations fulfilled by the $^1P_1$ and $^3P_1$ states, respectively, and then we mix their wave functions  to get those of the $1^{+}$ states.  

The Bethe-Salpeter wave function of $^1P_1$ is \cite{simi3}
\begin{eqnarray}
\begin{aligned}
\varphi^{++}=q_{\perp}\cdot\epsilon \left[ A_1(q_{\perp})+\frac{\slashed{P}}{M}A_2(q_{\perp})+\frac{\slashed{q}_{\perp}}{M}A_3(q_{\perp})+\frac{\slashed{P}\slashed{q}_{\perp}}{M^2}A_4(q_{\perp}) \right]\gamma_5,
\end{aligned}
\end{eqnarray}
where 
\begin{eqnarray}
\begin{aligned}
&A_1=\frac{1}{2}\left[ f_1+\frac{w_1+w_2}{m_1+m_2}f_2 \right],\\
&A_2=\frac{1}{2}\left[ \frac{m_1+m_2}{w_1+w_2}f_1+f_2 \right],\\
&A_3=-\frac{M(w_1-w_2)}{m_1w_2+m_2w_1}A_1,\\
&A_4=-\frac{M(m_1+m_2)}{m_1w_2+m_2w_1}A_1.
\end{aligned}
\end{eqnarray}
Here $M$ and $P$ are the mass and momentum of the initial meson; $q$ is the relative momentum between the quark and anti-quark in the initial meson; $q_\perp$ denotes $q- \frac{P\cdot q}{M}P$; and $m_1$, $m_2$ are the masses of the quark and anti-quark, respectively. The definition $w_i=\sqrt{m_i^2-q_\perp^2}(i=1,2)$ is used. $f_1$ and $f_2$ are the radial wave functions obtained by solving the Bethe-Salpeter equation.  

The Bethe-Salpeter wave function of $^3P_1$ is \cite{simi3}
\begin{eqnarray}
\begin{aligned}
\varphi^{++}=i\varepsilon_{\mu\nu\alpha\beta}\frac{P^{\nu}}{M}
q_{\perp}^{\alpha}\epsilon^{\beta}\gamma^{\mu} \left[ B_1(q_{\perp})+\frac{\slashed{P}}{M}B_2(q_{\perp})+\frac{\slashed{q}_{\perp}}{M}B_3(q_{\perp})+\frac{\slashed{P}\slashed{q}_{\perp}}{M^2}B_4(q_{\perp}) \right],
\end{aligned}
\end{eqnarray}
where
\begin{eqnarray}
\begin{aligned}
&B_1=\frac{1}{2}\left[ g_1+\frac{w_1+w_2}{m_1+m_2}g_2 \right],\\
&B_2=-\frac{1}{2}\left[ \frac{m_1+m_2}{w_1+w_2}g_1+g_2 \right],\\
&B_3=\frac{M(w_1-w_2)}{m_1w_2+m_2w_1}B_1,\\
&B_4=-\frac{M(m_1+m_2)}{m_1w_2+m_2w_1}B_1.
\end{aligned}
\end{eqnarray}

To get the $1^+$ state, we use the mixing equation \cite{simi3}
\begin{equation} \label{mixing}
\begin{aligned}
  &\left|\frac{3}{2} \right \rangle = {\cos{\theta}}\left|^1P_1\right\rangle+ {\sin {\theta}}\left |^3P_1\right\rangle,\\
  &\left|\frac{1}{2} \right \rangle = -{\sin{\theta}}\left|^1P_1\right\rangle +{\cos {\theta}} \left|^3P_1\right\rangle.
\end{aligned}
\end{equation}
In the heavy quark limit ($m_Q \rightarrow \infty$), the spin of the heavy quark $s_Q$ can be separated from the total angular momentum, so the heavy-light meson can be described by the good quantum number $j_l^P$, where $P$ is parity, and $\vec{j}_l=\vec{s}_q+\vec{L}$, with $\vec{s}_q$ and $\vec{L}$ denoting the spin of the light quark and the orbital angular momentum of the heavy-light meson, respectively. Thus, the 2P$(1^{+'})$ and 2P$(1^{+})$ states in the S doublet and T doublet can be denoted by $\left|\frac{1}{2} \right \rangle$ and $\left|\frac{3}{2} \right \rangle$, respectively. 

Apart from the mixing of wave functions, the mass mixing equation for two $1^+$ states is given as \cite{simi3}
\begin{eqnarray}
\begin{aligned}
&m^2_{^1P_1}=m^2_{1/2} \sin^2\theta +m^2_{3/2} \cos ^2 \theta,\\
&m^2_{^3P_1}=m^2_{1/2} \cos^2\theta +m^2_{3/2} \sin ^2 \theta.
\end{aligned}
\end{eqnarray}
In the equation, the masses of two physical states are needed, while we notice that the partners of $D_{sJ}(3040)^+$ and $D_J(3000)^0$ have not been discovered experimentally yet. Thus we adopt our theoretical mass predictions of the two partners. Table I shows the masses in our model and in other models as well. 
\begin{center}
\begin{table*}[h]
\caption{Mass spectrum of the 2P states in the $D$ and $D_s$ families (in units of MeV).}
\begin{center}
\begin{tabular}{ccccccc}
\hline\hline
\rule[-2mm]{0mm}{6.5mm}
State       &\qquad {ours} &\quad Ref.~\cite{mass8} & \quad Ref.~\cite{mass9} &\quad Ref.~\cite{mass11} & \quad  Ref.~\cite{mass12} \\
\hline
\rule[-2mm]{0mm}{6.5mm}
$D(2^1P_1)$ &\qquad 2933  & \quad 2940  & \quad2932 & \quad3045 & \\
\rule[-2mm]{0mm}{6.5mm}
$D(2^3P_1)$ &\qquad 2952  &  \quad 2960  & \quad3021 & \quad2995 &  \\
\rule[-2mm]{0mm}{6.5mm}
$D_s(2^1P_1)$ &\qquad 3029 & \quad 3040  &\quad 3067 &  \quad3165  & \quad 2959.0\\
\rule[-2mm]{0mm}{6.5mm}
$D_s(2^3P_1)$ &\qquad 3036 & \quad 3020 &  \quad3154  & \quad 3114  & \quad 2986.4 \\
 \hline\hline
\end{tabular}
\end{center}
\label{2Pmassspectrum} 
\end{table*}
\end{center}

If both the initial and final mesons are $1^+$ states, for example, in the $D_{sJ}(3040)^+\rightarrow D_1(2420)^0K^+$ channel, the mixing matrix of the amplitude will be the direct product of the mixing matrices of the wave functions, which is  a $4 \times 4$ matrix. The mixing equation in this case takes the form of \cite{wangzhigang3000}
\begin{gather}
\begin{bmatrix} \mathcal {M}_{1^{+'}\rightarrow1^{+'}}  \\ \mathcal {M}_{1^{+'}\rightarrow1^{+}}\\ \mathcal {M}_{1^{+}\rightarrow1^{+'}}\\\mathcal {M}_{1^{+}\rightarrow1^{+}}\end{bmatrix}
=\begin{bmatrix} \cos{\theta}\cos{\theta}' && -\sin{\theta}\cos{\theta}' &&-\cos{\theta}\sin{\theta}' &&\sin{\theta}\sin{\theta}' \\ \cos{\theta}\sin{\theta}' && -\sin{\theta} \sin{\theta}' && \cos{\theta}\cos{\theta}' && -\sin{\theta}\cos{\theta}' \\  \sin{\theta}\cos{\theta}' && \cos{\theta}\cos{\theta}' &&-\sin{\theta}\sin{\theta}' &&-\cos{\theta}\sin{\theta}' \\ \sin{\theta}\sin{\theta}'&& \cos{\theta}\sin{\theta}'&& \sin{\theta}\cos{\theta}'&& \cos{\theta}\cos{\theta}'\end{bmatrix}    
\begin{bmatrix} \mathcal {M}_{^3P_1 \rightarrow ^3P_1}  \\ \mathcal {M}_{^1P_1 \rightarrow ^3P_1}  \\ \mathcal {M}_{^3P_1 \rightarrow ^1P_1}  \\ \mathcal {M}_{^1P_1 \rightarrow ^1P_1}  \end{bmatrix}.
\end{gather}

For the final mesons, the quantum numbers include $0^-$, $0^+$, $1^-$, $1^+$, $1^{+'}$, $2^+$. We take the $1^-$ state as an example. Other states can be found in our previous works \cite{wangzhihuiwavefunction0,wangguoliwavefunction2}. The Bethe-Salpeter wave function of the $1^-$ state is
\begin{eqnarray}
\begin{aligned}
\varphi_{1^-}^{++}&=q_{\perp}\cdot\epsilon \left[ C_1(q_{\perp})+\frac{\slashed{P}}{M}C_2(q_{\perp})+\frac{\slashed{q}_{\perp}}{M}C_3(q_{\perp})+\frac{\slashed{P}\slashed{q}_{\perp}}{M^2}C_4(q_{\perp}) \right] \\
               &+M\slashed{\epsilon}\left[ C_5(q_{\perp})+\frac{\slashed{P}}{M}C_6(q_{\perp})+\frac{\slashed{q}_{\perp}}{M}C_7(q_{\perp})+\frac{\slashed{P}\slashed{q}_{\perp}}{M^2}C_8(q_{\perp}) \right],
\end{aligned}
\end{eqnarray}
where 
\begin{eqnarray}
\begin{aligned}
&C_1=\frac{1}{2M(m_1w_2+m_2w_1)} \left[ (w_1+w_2)q^2_{\perp}f_3 + (m_1+m_2)q^2_{\perp}f_4 + 2M^2w_2f_5 - 2M^2m_2f_6  \right],\\
&C_2=\frac{1}{2M(m_1w_2+m_2w_1)} \left[ (m_1-m_2)q^2_{\perp}f_3 + (w_1-w_2)q^2_{\perp}f_4 - 2M^2m_2f_5 + 2M^2w_2f_6  \right],\\
&C_3=\frac{1}{2} \left[ f_3+\frac{m_1+m_2}{w_1+w_2}f_4-\frac{2M^2}{m_1w_2+m_2w_1}f_6 \right], ~~~~C_5=\frac{1}{2}  \left[ f_5-\frac{w_1+w_2}{m_1+m_2}f_6 \right],\\
&C_4=\frac{1}{2} \left[ \frac{w_1+w_2}{m_1+m_2}f_3+f_4-\frac{2M^2}{m_1w_2+m_2w_1}f_5 \right],~~~~C_6=\frac{1}{2}  \left[ -\frac{m_1+m_2}{w_1+w_2}f_5+f_6 \right],\\
&C_7=\frac{M}{2} \frac{w_1-w_2}{m_1w_2+m_2w_1} \left[ f_5-\frac{w_1+w_2}{m_1+m_2}f_6 \right],\quad\quad\quad\quad\quad\quad\quad\quad\quad\quad\quad\quad\quad\quad\quad\quad\quad\quad\\
&C_8=\frac{M}{2} \frac{m_1+m_2}{m_1w_2+m_2w_1} \left[ -f_5+\frac{w_1+w_2}{m_1+m_2}f_6 \right].
\end{aligned}
\end{eqnarray}

For the final state, the wave function should take the Dirac conjugate form, which is $\overline{\varphi}_{1^-}^{++}=\gamma_0 (\varphi_{1^-}^{++})^+ \gamma_0$ for mesons.
In the calculation, the completeness relations fulfilled by the polarization vector (tensor) are applied, which read as
\begin{eqnarray}
\begin{aligned}
\sum_{r}\epsilon_{(r)}^{\mu}\epsilon_{(r)}^{*\nu}=&-g^{\mu\nu}+\frac{P^{\mu}P^{\nu}}{M^2},\\
\sum_{r}\epsilon_{(r)}^{\mu\nu}\epsilon_{(r)}^{*\alpha\beta}=&\frac{1}{2} \left[ \left( -g^{\mu\alpha}+\frac{P^{\mu}P^{\alpha}}{M^2}\right) \left( -g^{\nu\beta}+\frac{P^{\nu}P^{\beta}}{M^2}\right) + \left( -g^{\mu\beta}+\frac{P^{\mu}P^{\beta}}{M^2}\right) \right.\\
& \left. \left( -g^{\nu\alpha}+\frac{P^{\nu}P^{\alpha}}{M^2} \right) \right]  -\frac{1}{3}\left( -g^{\mu\nu}+\frac{P^{\mu}P^{\nu}}{M^2}\right)\left( -g^{\alpha\beta}+\frac{P^{\alpha}P^{\beta}}{M^2}\right),
\end{aligned}
\end{eqnarray}
where the polarization vector satisfies $\epsilon\cdot P= 0$, and the polarization tensor satisfies $\epsilon^{\mu\nu}P_{\mu}=0$, $\epsilon^{\mu\nu}g_{\mu\nu}=0$.

\section{NUMERICAL RESULTS AND DISCUSSIONS}

In this section, we give our results and compare ours with those from other models. In our model, the parameters are set as follows: $m_u=0.305\;\rm{GeV}$, $m_d=0.311\;\rm{GeV}$,  $m_c=1.620\;\rm{GeV}$, $m_s=0.500\;\rm{GeV}$, and $m_b=4.960\;\rm{GeV}$. For the masses of the partners of $D_J(3000)$ and $D_{sJ}(3040)$, which have not been discovered yet, we choose our theoretical predictions: $D_s(2\rm{P}1^{+})=3.022\;\rm{GeV}$ and $D(2\rm{P}1^{+})=2.975\;\rm{GeV}$; $m_{\eta_0}=0.923\;\rm{GeV}$ and $m_{\eta_8}=0.604\;\rm{GeV}$; and decay constants $f_{\pi}=0.1304\;\rm{GeV}$, $f_K=0.1562\;\rm{GeV}$ \cite{PDG}, $f_{\eta_0}=1.07f_{\pi}$, and $f_{\eta_8}=1.26f_{\pi}$.

\begin{figure}[htbp]
\centering
\subfigure[$^1P_1$wavefunction \label{$^1P_1$wavefunction}]
{\includegraphics[width=0.47\textwidth]{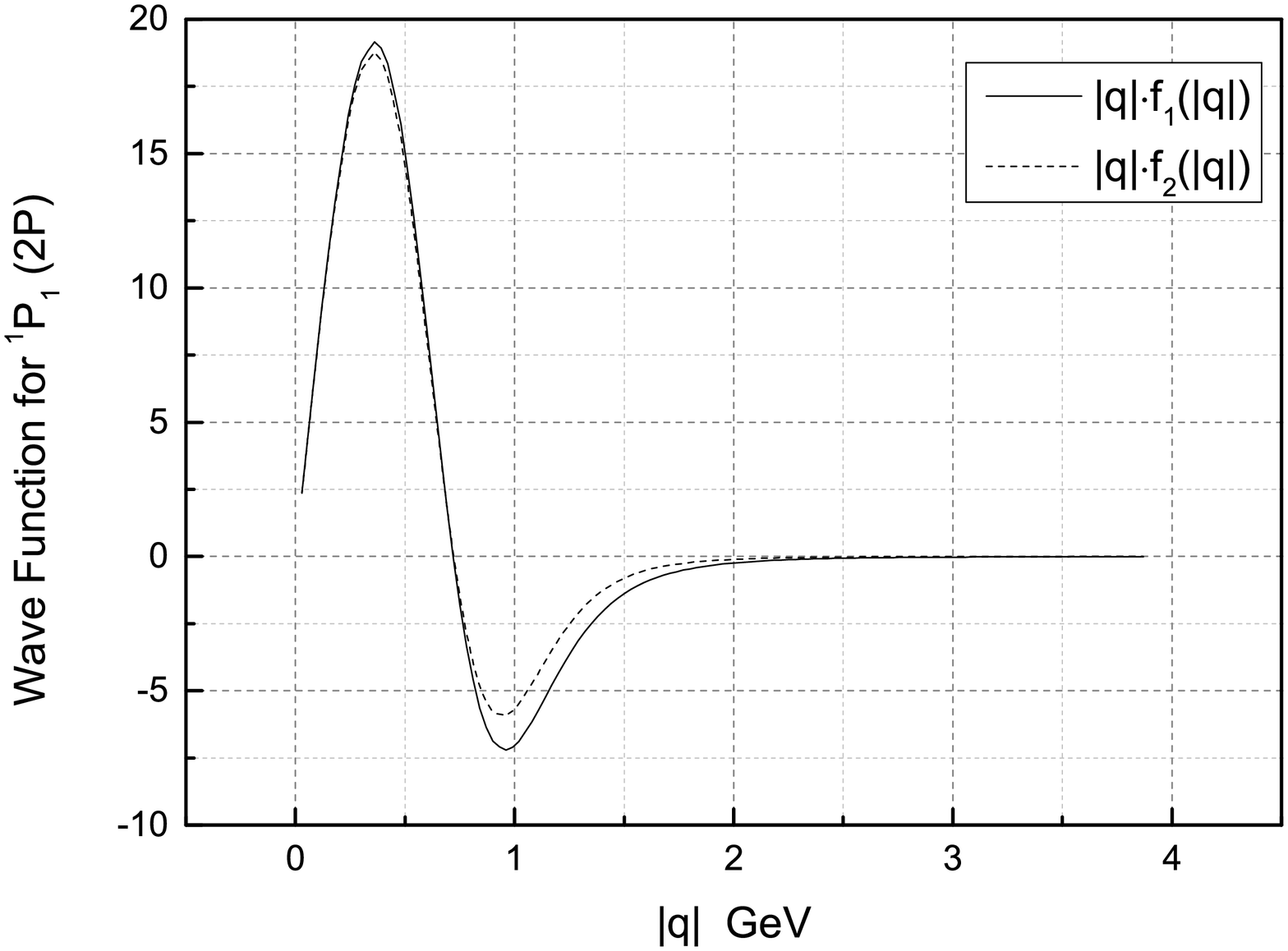}}
\subfigure[$^3P_1$wavefunction\label{$^3P_1$wavefunction}]
{\includegraphics[width=0.47\textwidth]{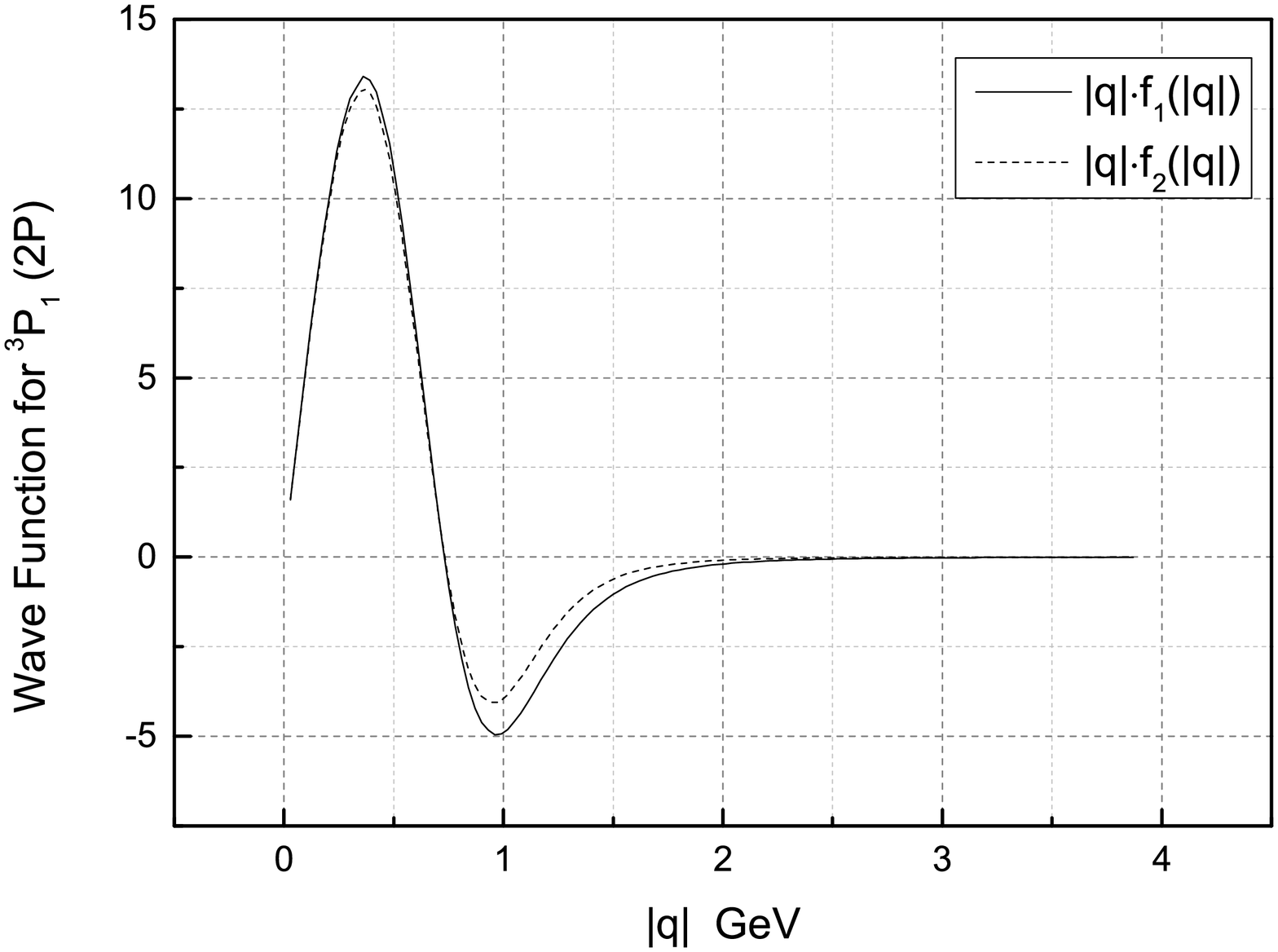}}
\caption{(a) The wave function for $^1P_1(2{\rm P})$ state (b) The wave function for $^3P_1(2{\rm P})$ state.}\label{wavefunction}
\vspace{-1em}
\end{figure}

The wave functions of the initial and final meson could be obtained by solving the instantaneous Bethe-Salpeter equation. In this process, we choose the Cornell potential and the explicit form could be found in Ref. \cite{cornellpotential}. We take the wave functions of $D_{sJ}(3040)$ as an example, which are shown in Fig. \ref{wavefunction}

\subsection{For $D_J(3000)$}

Tables.~II and III show the decay widths of $D_J(3000)^0$ as 2P$(1^{+'})$ and 2P$(1^{+})$ states, respectively. In order to show the relative values, we give the branching ratios of different channels under the assumption of the 2P$(1^{+'})$ and 2P$(1^{+})$ states in Tables.~IV and V. In these tables, ``$-$'' denotes the  forbidden channel. ``$\square$''  denotes the channel is allowed but not calculated in the corresponding literature. $D^*(2600)$ and $D^*(2650)$ are treated as pure $2^3S_1$ and $1^3D_1$ states, respectively. The mixing angle between $\phi_8$ and $\phi_0$ is $19 ^\circ$ in this paper. If we choose the result of ChPT in the tree level, it is $9.95^\circ$. This factor will affect little to the result. For example, the two largest channels involving $\eta$ or $\eta^\prime$ are $ D^*(2007)^0\eta$ and $D^*(2007)^0\eta'$, with partial widths 5.03 MeV and 4.15 MeV, respectively. If we use a mixing angle of $9.95^\circ$ in these channels, then the partial widths are 5.97 MeV and 7.91 MeV correspondingly.  

The first thing we notice in Tables.~II and III is the total width. Our result is larger than the central value but less than the upper limit of 232.9 MeV of the experiment under the assumption of the 2P$(1^{+'})$ state. Under the assumption of the 2P$(1^{+})$ state, it will be much less than the lower limit of the experiment. Another comparison with experimental data is about the dominant channel. $D_J(3000)$ was first observed in the $D^*\pi$ channel. In our calculation, $D^*\pi$ and $D^*(2600)\pi$ share almost the same proportion just next to $D^*_2(2460)\pi$ for the 2P$(1^{+'})$ state, but the $D^*\pi$ channel is ignorable under the 2P$(1^{+})$ assignment. Therefore, $D_J(3000)^0$ is a good candidate for the 2P$(1^{+'})$ state. Because very little information has been given in experiments, we recommend the detection of the channels of $D^*_2(2460)\pi$, $D^*\pi$, and $D^*(2600)\pi$. These channels are dominant channels in our results, and the precise detection of them can help us to distinguish the quantum states from the 2P$(1^+)$ state. Moreover, the ratio of the partial widths of $D^*_2(2460)\pi$, $D^*(2600)\pi$, and $D^*\pi$ is $1:0.44:0.44$ in our calculation, which can also be used in comparison with future experimental results.

In Ref. \cite{liuxiang3000}, Liu $et$ $al.$ employed a QPC model to give similar results in most channels but smaller than ours in the  $D^*_2(2460)\pi$ channel. Some allowed channels such as $D^*(2600)\pi$ and $D^*(2650)\pi$ were not calculated in their work. These missing modes may contribute to the total width difference. Besides the QPC model, Liu $et$ $al.$ also use the modified Godfrey-Isgur (G-I) model to calculate the same channels in Ref. \cite{liuxiang3000modifiedGI}, while the results in this model are 289.41 MeV for the 2P$(1^{+'})$ state and 97.31 MeV for the 2P$(1^{+})$ state. In Ref. \cite{Godfreyresults}, decay widths of some dominant channels were calculated by Godfrey $et$ $al.$ in the G-I model. Under the assumption of the 2P$(1^{+'})$ state, the largest channel is $D^*_2(2460)\pi$ in their result, sharing the same values with ours at around 80 MeV. One thing should be mentioned is about the mass. The mass used in that work was 2961 MeV, which is in their theoretical prediction, rather than the 2971 MeV mass used in experiments, but this leads to a very little difference. In the results of Ref. \cite{lidemin3000}, both 2P$(1^{+'})$ and 2P$(1^{+})$ states are larger than the experimental data, so the authors excluded these quantum states. In Ref. \cite{wangzhigang3000}, the total width given by Wang $et$ $al.$ is very close to the experimental data, but the width of 2P$(1^{+'})$ is less than that of 2P$(1^{+})$  in their results, which is different to our knowledge. In addition, in Ref. \cite{ZhongXianHui3000}, the partial and total decay widths as functions of the mass and the mixing angle were given. With a mixing angle of $-54.7^{\circ}$ derived in the heavy quark limit, the total width was around 360 MeV for the 2P$(1^{+'})$ state, which is about two times larger than the experimental data. In Ref. \cite{effctive3000_revise_add}, the authors used the effective Lagrangian method to give an analysis for  some dominant strong decays; they also favor $D_J(3000)$ as the 2P$(1^{+'})$ state.

\begin{table}
\begin{center}
\linespread{0.5}
\begin{longtable}{ccccccc}
\caption[]{The partial and total widths (in units of MeV) of $D_J(3000)^0$ as the 2P$(1^{+'})$ state.} \label{} \\
\hline\hline
\rule[-2mm]{0mm}{6.5mm}{}&{Final state}&{\bfseries Ours}&{\bfseries Ref.\cite{liuxiang3000}}&{\bfseries Ref.\cite{wangzhigang3000}}&{\bfseries Ref.\cite{Godfreyresults}}&{\bfseries Ref. \cite{lidemin3000}}\\
\hline
\rule[-2mm]{0mm}{6.5mm}
$1^+\rightarrow1^-0^-$ & $ D^*(2007)^0\pi^0$ &  $13.37$ & $\multirow{2}{*}{38}$ &  $10.03$ &\multirow{2}{*}{21.6}&18.79\\
\rule[-2mm]{0mm}{6.5mm}
$$&$ D^*(2010)^+\pi^-$ & $25.40$ & $$ &  $20.32$&&36.92\\
\rule[-2mm]{0mm}{6.5mm}
$$&$ D^*(2007)^0\eta$ & $5.03$ & $5.2$&  $4.92$&$\square$&4.39\\
\rule[-2mm]{0mm}{6.5mm}
$$&$ D^*(2007)^0\eta'$ & $4.15$ & $0.023$& $2.71$&$\square$&3.80\\
\rule[-2mm]{0mm}{6.5mm}
$$ & $ D^*(2600)^0\pi^0$ & $13.14$ & $\square$ & $\square$ &\multirow{2}{*}{20.9}&20.90 \\
\rule[-2mm]{0mm}{6.5mm}
$$ & $ D^*(2600)^+\pi^-$ & $26.28$ & $\square$ & $\square$ & &42.04\\
\rule[-2mm]{0mm}{6.5mm}
$$ & $ D^*(2650)^0\pi^0$ & $1.01$ & $\square$ & $\square$ & $\square$ &0.02\\
\rule[-2mm]{0mm}{6.5mm}
$$ & $ D^*(2650)^+\pi^-$ & $2.02$ & $\square$ & $\square$ & $\square$ &0.32\\
\rule[-2mm]{0mm}{6.5mm}
$1^+\rightarrow0^-1^-$ & $ D^{0}\rho^{0}$ & $2.00$ & $\multirow{2}{*}{7.6}$& $5.61$&\multirow{2}{*}{18.8}&26.99\\
\rule[-2mm]{0mm}{6.5mm}
$$&$ D^{+}\rho^{-}$ & $4.26$ &$$& $10.59$&&53.14\\
\rule[-2mm]{0mm}{6.5mm}
$$&$ D^{0}\omega$ & $2.15$ & $2.5$ & $4.99$&6.11&26.55\\
\rule[-2mm]{0mm}{6.5mm}
$1^+\rightarrow1^-1^-$ & $ D^{*}(2007)^0\rho^{0}$ & $5.51$ & $\multirow{2}{*}{15}$ & $21.07$&\multirow{2}{*}{23.3}&29.47\\
\rule[-2mm]{0mm}{6.5mm}
$$&$ D^{*}(2010)^+\rho^{-}$ & $10.38$ &$$& $41.34$&&57.33\\
\rule[-2mm]{0mm}{6.5mm}
$$&$ D^{*}(2007)^0\omega$ & $5.41$ & $4.9$ & $19.93$&7.3&28.70\\
\rule[-2mm]{0mm}{6.5mm}
$1^+\rightarrow0^+0^-$ & $ D^*_0(2400)^{0}\pi^{0}$ & $1.90$ & $\multirow{2}{*}{6}$ & $0.24$&$\square$&1.93\\
\rule[-2mm]{0mm}{6.5mm}
$$&$ D^*_0(2400)^{+}\pi^{-}$ & $4.09$ & $$ & $\square$&$\square$&4.06\\
\rule[-2mm]{0mm}{6.5mm}
$$&$ D^*_0(2400)^{0}\eta$ & $0.53$ & $0.068$ & $0.27$&$\square$&0.84\\
\rule[-2mm]{0mm}{6.5mm}
$1^+\rightarrow1^+0^-$ & $ D_1(2420)^0\pi^0$ & $2.33$ & $\multirow{2}{*}{14}$ & $0.0081$&\multirow{2}{*}{15.9}&2.77\\
\rule[-2mm]{0mm}{6.5mm}  
$$ & $ D_1(2420)^+\pi^-$ & $4.69$ & $$ & $\square$&&5.53\\
\rule[-2mm]{0mm}{6.5mm}
$$&$ D_1(2420)\eta$ & $0.0023$ & $0.0042$ & $0.003$&$\square$&0.0072\\
\rule[-2mm]{0mm}{6.5mm}
$$&$ D_1(2430)^0\pi^0$ & $1.84$ &$\multirow{2}{*}{11}$ & $0.0099$&\multirow{2}{*}{5.3}&0.11\\
\rule[-2mm]{0mm}{6.5mm}
$$&$ D_1(2430)^+\pi^-$ & $3.64$ &$$ & $\square$&&0.21\\
\rule[-2mm]{0mm}{6.5mm}
$$&$ D_1(2430)^0\eta$ & $-$ &$-$ & $0.0015$&$-$&$-$\\
\rule[-2mm]{0mm}{6.5mm}
$1^+\rightarrow2^+0^-$ & $ D_2^*(2460)^{0}\pi^{0}$ & $30.69$ & $\multirow{2}{*}{38}$ & $5.39$&\multirow{2}{*}{82.3}&40.40\\
\rule[-2mm]{0mm}{6.5mm}
$$&$ D_2^*(2460)^{+}\pi^{-}$ & $58.01$ & $$ & $10.52$&&80.53\\
\rule[-2mm]{0mm}{6.5mm}
$$&$ D_2^*(2460)^{0}\eta$ & $-$ & $-$ & $0.024$&$-$&$-$\\
\hline
\rule[-2mm]{0mm}{6.5mm}
$1^+\rightarrow0^-1^-$ & $ D_s^{+}K^{*-}$ & $0.12$ & $0.12$ & $7.13$&4.0&1.48\\
\rule[-2mm]{0mm}{6.5mm}
$1^+\rightarrow1^-0^-$ & $ D_s^{*+}K^-$ & $1.14$ & $3.7$ & $9.45$&4.4&0.95\\
\rule[-2mm]{0mm}{6.5mm}
$1^+\rightarrow0^+0^-$ & $ D^*_{s0}(2317)^{+}K^{-}$ & $0.42$ & $0.67$ & $0.83$&$\square$&1.19\\
\rule[-2mm]{0mm}{6.5mm}
$1^+\rightarrow1^-1^-$ & $ D_s^{*+}K^{*}$ & $-$ & $-$ & $2.05$&$-$&$-$\\
\rule[-2mm]{0mm}{6.5mm}
$1^+\rightarrow1^+0^-$ & $ D_{s1}(2460)^{+}K^{-}$ & $0.049$ & $0.082$ & $0.0081$&$\square$&0.00021\\
\rule[-2mm]{0mm}{6.5mm}
$$ & $ D_{s1}(2536)^{+}K^{-}$ & $-$ & $-$ & $0.024$&$-$&$-$\\
\hline
\rule[-2mm]{0mm}{6.5mm}
{\rm Total} & $Exp:188.1\pm 44.8$ & $229.6$ & $146.8$ & $177.5$&209.9&489.3\\
 \hline\hline
\end{longtable}

\end{center}
\end{table}

\begin{table}
\begin{center}
\linespread{0.5}
\begin{longtable}{ccccccc}
\caption[]{Decay widths (in units of MeV) of $D_J(3000)^0$ as the 2P$(1^{+})$ state.} \label{} \\
\hline\hline
\rule[-2mm]{0mm}{6.5mm}{}&{Final state}&{\bfseries Ours}&{\bfseries Ref.\cite{liuxiang3000}}&{\bfseries Ref.\cite{wangzhigang3000}}&{\bfseries Ref.\cite{Godfreyresults}}&{\bfseries Ref. \cite{lidemin3000}}\\
\hline
\rule[-2mm]{0mm}{6.5mm}
$1^+\rightarrow1^-0^-$ & $ D^*(2007)^0\pi^0$ &  $0.97$ & $\multirow{2}{*}{1.3}$ &  $11.85$ &\multirow{2}{*}{37.9}&$15.64$\\
\rule[-2mm]{0mm}{6.5mm}
$$&$ D^*(2010)^+\pi^-$ & $1.83$ & $$ &  $23.62$&&$31.25$\\
\rule[-2mm]{0mm}{6.5mm}
$$&$ D^*(2007)^0\eta$ & $0.10$ & $0.49$&  $2.48$&$5.0$&$6.88$\\
\rule[-2mm]{0mm}{6.5mm}
$$&$ D^*(2007)^0\eta'$ & $0.08$ & $0.00026$& $18.72$&$\square$&$0.95$\\
\rule[-2mm]{0mm}{6.5mm}
$$ & $ D^*(2600)^0\pi^0$ & $4.78$ & $\square$ & $\square$ &\multirow{2}{*}{1.3}&$5.93$ \\
\rule[-2mm]{0mm}{6.5mm}
$$ & $ D^*(2600)^+\pi^-$ & $9.56$ & $\square$ & $\square$ & &$11.84$\\
\rule[-2mm]{0mm}{6.5mm}
$$ & $ D^*(2650)^0\pi^0$ & $0.26$ & $\square$ & $\square$ & $\square$ &$0.02$\\
\rule[-2mm]{0mm}{6.5mm}
$$ & $ D^*(2650)^+\pi^-$ & $0.52$ & $\square$ & $\square$ & $\square$ &$0.04$\\
\rule[-2mm]{0mm}{6.5mm}
$1^+\rightarrow0^-1^-$ & $ D^{0}\rho^{0}$ & $1.90$ & $\multirow{2}{*}{4.7}$& $17.27$&\multirow{2}{*}{3.4}&$1.16$\\
\rule[-2mm]{0mm}{6.5mm}
$$&$ D^{+}\rho^{-}$ & $4.31$ &$$& $34.52$&&$1.98$\\
\rule[-2mm]{0mm}{6.5mm}
$$&$ D^{0}\omega$ & $1.89$ & $1.5$ & $17.30$&1.1&$0.95$\\
\rule[-2mm]{0mm}{6.5mm}
$1^+\rightarrow1^-1^-$ & $ D^{*}(2007)^0\rho^{0}$ & $11.30$ & $\multirow{2}{*}{14}$ & $18.46$&\multirow{2}{*}{24.4}&$32.84$\\
\rule[-2mm]{0mm}{6.5mm}
$$&$ D^{*}(2010)^+\rho^{-}$ & $21.29$ &$$& $36.26$&&$62.68$\\
\rule[-2mm]{0mm}{6.5mm}
$$&$ D^{*}(2007)^0\omega$ & $10.90$ & $4.6$ & $17.53$&8.2&$31.31$\\
\rule[-2mm]{0mm}{6.5mm}
$1^+\rightarrow0^+0^-$ & $ D^*_0(2400)^{0}\pi^{0}$ & $0.46$ & $\multirow{2}{*}{11}$ & $0.17$&$\multirow{2}{*}{4.9}$&$0.94$\\
\rule[-2mm]{0mm}{6.5mm}
$$&$ D^*_0(2400)^{+}\pi^{-}$ & $1.40$ & $$ & $\square$&$$&$1.98$\\
\rule[-2mm]{0mm}{6.5mm}
$$&$ D^*_0(2400)^{0}\eta$ & $0.23$ & $0.14$ & $0.30$&$\square$&$0.4$\\
\rule[-2mm]{0mm}{6.5mm}
$1^+\rightarrow1^+0^-$ & $ D_1(2420)^0\pi^0$ & $2.34$ & $\multirow{2}{*}{8.8}$ & $0.024$&\multirow{2}{*}{5.2}&$11.32$\\
\rule[-2mm]{0mm}{6.5mm}  
$$ & $ D_1(2420)^+\pi^-$ & $4.70$ & $$ & $\square$&&$22.62$\\
\rule[-2mm]{0mm}{6.5mm}
$$&$ D_1(2420)\eta$ & $0.0029$ & $0.0023$ & $0.0061$&$\square$&$0.03$\\
\rule[-2mm]{0mm}{6.5mm}
$$&$ D_1(2430)^0\pi^0$ & $0.15$ &$\multirow{2}{*}{5.3}$ & $0.0081$&\multirow{2}{*}{2.5}&$0.15$\\
\rule[-2mm]{0mm}{6.5mm}
$$&$ D_1(2430)^+\pi^-$ & $0.30$ &$$ & $\square$&&$0.29$\\
\rule[-2mm]{0mm}{6.5mm}
$$&$ D_1(2430)^0\eta$ & $-$ &$-$ & $0.003$&$-$&$-$\\
\rule[-2mm]{0mm}{6.5mm}
$1^+\rightarrow2^+0^-$ & $ D_2^*(2460)^{0}\pi^{0}$ & $2.89$ & $\multirow{2}{*}{3.3}$ & $28.05$&\multirow{2}{*}{7.4}&$6.98$\\
\rule[-2mm]{0mm}{6.5mm}
$$&$ D_2^*(2460)^{+}\pi^{-}$ & $5.68$ & $$ & $56.21$&&$13.70$\\
\rule[-2mm]{0mm}{6.5mm}
$$&$ D_2^*(2460)^{0}\eta$ & $-$ & $-$ & $0.56$&$-$&$-$\\
\rule[-2mm]{0mm}{6.5mm}
$1^+\rightarrow0^-1^-$ & $ D_s^{+}K^{*-}$ & $0.41$ & $0.7$ & $3.82$&14.3&$10.41$\\
\rule[-2mm]{0mm}{6.5mm}
$1^+\rightarrow1^-0^-$ & $ D_s^{*+}K^-$ & $0.055$ & $0.099$ & $1.22$&9.0&$4.14$\\
\rule[-2mm]{0mm}{6.5mm}
$1^+\rightarrow0^+0^-$ & $ D^*_{s0}(2317)^{+}K^{-}$ & $5.29$ & $1.2$ & $0.52$&$\square$&$0.74$\\
\rule[-2mm]{0mm}{6.5mm}
$1^+\rightarrow1^-1^-$ & $ D_s^{*+}K^{*}$ & $-$ & $-$ & $4.08$&$-$&$-$\\
\rule[-2mm]{0mm}{6.5mm}
$1^+\rightarrow1^+0^-$ & $ D_{s1}(2460)^{+}K^{-}$ & $0.043$ & $0.045$ & $0.024$&$\square$&$0.01$\\
\rule[-2mm]{0mm}{6.5mm}
$$ & $ D_{s1}(2536)^{+}K^{-}$ & $-$ & $-$ & $0.049$ & $-$ & $-$\\
\hline
\rule[-2mm]{0mm}{6.5mm}
{\rm Total} & $Exp:188.1\pm 44.8$ & $93.6$ & $57.1$ & $293.1$&124.6&277.2\\
 \hline\hline
\end{longtable}
\end{center}
\end{table}

\begin{table}
\begin{center}
\linespread{0.5}
\begin{longtable}{ccccccc}
\caption[]{Branching ratios of different decay channels of $D_J(3000)^0$ as the 2P$(1^{+'})$ state.} \label{} \\
\hline\hline
\rule[-2mm]{0mm}{6.5mm}{}&{Final state}&{\bfseries Ours}&{\bfseries Ref.\cite{liuxiang3000}}&{\bfseries Ref.\cite{wangzhigang3000}}&{\bfseries Ref.\cite{Godfreyresults}}&{\bfseries Ref. \cite{lidemin3000}}\\
\hline
\rule[-2mm]{0mm}{6.5mm}
$1^+\rightarrow1^-0^-$ & $ D^*(2007)^0\pi^0$ &  $5.82\%$ & $\multirow{2}{*}{25.87\%}$ &  $5.65\%$& $\multirow{2}{*}{10.2\%}$&3.83\%\\
\rule[-2mm]{0mm}{6.5mm}
$$&$ D^*(2010)^+\pi^-$ & $11.06\% $ & $$ &  $11.45\%$&&7.54\%\\
\rule[-2mm]{0mm}{6.5mm}
$$&$ D^*(2007)^0\eta$ & $2.19\%$ & $3.54\%$&  $2.78\%$& $\square$&0.89\%\\
\rule[-2mm]{0mm}{6.5mm}
$$&$ D^*(2007)^0\eta'$ & $1.81\%$ & $0.02\%$& $1.53\%$& $\square$&0.77\%\\
\rule[-2mm]{0mm}{6.5mm}
$$ & $ D^*(2600)^0\pi^0$ & $5.72\%$ & $\square$ & $\square$ &\multirow{2}{*}{9.96\%}&4.27\% \\
\rule[-2mm]{0mm}{6.5mm}
$$ & $ D^*(2600)^+\pi^-$ & $11.45\%$ & $\square$ & $\square$ & &8.59\%\\
\rule[-2mm]{0mm}{6.5mm}
$$ & $ D^*(2650)^0\pi^0$ & $0.44\%$ & $\square$ & $\square$ & $\square$ &0.004\%\\
\rule[-2mm]{0mm}{6.5mm}
$$ & $ D^*(2650)^+\pi^-$ & $0.88\%$ & $\square$ & $\square$ & $\square$ &0.06\%\\
\rule[-2mm]{0mm}{6.5mm}
$1^+\rightarrow0^-1^-$ & $ D^{0}\rho^{0}$ & $0.87\%$ & $\multirow{2}{*}{5.17\%}$& $3.16\%$&$\multirow{2}{*}{8.9\%}$&5.51\%\\
\rule[-2mm]{0mm}{6.5mm}
$$&$ D^{+}\rho^{-}$ & $1.86\%$ &$$& $5.97\%$&&10.86\%\\
\rule[-2mm]{0mm}{6.5mm}
$$&$ D^{0}\omega$ & $0.94\%$ & $1.70\%$ & $2.81\%$&2.9\%&5.42\%\\
\rule[-2mm]{0mm}{6.5mm}
$1^+\rightarrow1^-1^-$ & $ D^{*}(2007)^0\rho^{0}$ & $2.40\%$ & $\multirow{2}{*}{10.21\%}$ & $11.87\%$&$\multirow{2}{*}{11.0\%}$&6.02\%\\
\rule[-2mm]{0mm}{6.5mm}
$$&$ D^{*}(2010)^+\rho^{-}$ & $4.52\%$ &$$& $23.29\%$&&11.72\%\\
\rule[-2mm]{0mm}{6.5mm}
$$&$ D^{*}(2007)^0\omega$ & $2.36\%$ & $3.33\%$ & $11.23\%$&3.5 \%&5.86\% \\
\rule[-2mm]{0mm}{6.5mm}
$1^+\rightarrow0^+0^-$ & $ D^*_0(2400)^{0}\pi^{0}$ & $0.83\%$ & $\multirow{2}{*}{4.09\%}$ & $\multirow{2}{*}{0.14\%}$ &$\square$&0.39\%\\
\rule[-2mm]{0mm}{6.5mm}
$$&$ D^*_0(2400)^{+}\pi^{-}$ & $1.78\%$ & $$ & $$ &$\square$&0.83\%\\
\rule[-2mm]{0mm}{6.5mm}
$$&$ D^*_0(2400)^{0}\eta$ & $0.23\%$ & $0.05\%$ & $0.15\%$&$\square$&0.17\%\\
\rule[-2mm]{0mm}{6.5mm}
$1^+\rightarrow1^+0^-$ & $ D_1(2420)^0\pi^0$ & $1.01\%$ & $\multirow{2}{*}{9.53\%}$ & $\multirow{2}{*}{0.004\%}$&$\multirow{2}{*}{7.5\%}$&0.57\%\\
\rule[-2mm]{0mm}{6.5mm}  
$$ & $ D_1(2420)^+\pi^-$ & $2.04\%$ & $$ & $$&&1.13\%\\
\rule[-2mm]{0mm}{6.5mm}
$$&$ D_1(2420)\eta$ & $0.001\%$ & $0.002\%$ & $0.002\%$&$\square$&0.001\%\\
\rule[-2mm]{0mm}{6.5mm}
$$&$ D_1(2430)^0\pi^0$ & $0.80\%$ &$\multirow{2}{*}{7.49\%}$ & $\multirow{2}{*}{0.005\%}$&$\multirow{2}{*}{2.5\%}$&0.02\%\\
\rule[-2mm]{0mm}{6.5mm}
$$&$ D_1(2430)^+\pi^-$ & $1.58\%$ &$$ & $$&&0.04\%\\
\rule[-2mm]{0mm}{6.5mm}
$$&$ D_1(2430)^0\eta$ & $-$ &$-$ & $0.0008\%$&$-$&$-$\\
\rule[-2mm]{0mm}{6.5mm}
$1^+\rightarrow2^+0^-$ & $ D_2^*(2460)^{0}\pi^{0}$ & $13.37\%$ & $\multirow{2}{*}{25.87\%}$ & $3.03\%$&$\multirow{2}{*}{38.9\%}$&8.26\%\\
\rule[-2mm]{0mm}{6.5mm}
$$&$ D_2^*(2460)^{+}\pi^{-}$ & $25.27\%$ & $$ & $5.93\%$&&16.45\%\\
\rule[-2mm]{0mm}{6.5mm}
$$&$ D_2^*(2460)^{0}\eta$ & $-$ & $-$ & $0.01\%$&$-$&$-$\\
\hline
\rule[-2mm]{0mm}{6.5mm}
$1^+\rightarrow0^-1^-$ & $ D_s^{+}K^{*-}$ & $0.05\%$ & $0.09\%$ & $4.01\%$& 1.9\%&0.30\% \\
\rule[-2mm]{0mm}{6.5mm}
$1^+\rightarrow1^-0^-$ & $ D_s^{*+}K^-$ & $0.50\%$ & $2.52\%$ & $5.32\%$&2.09\%&0.19\% \\
\rule[-2mm]{0mm}{6.5mm}
$1^+\rightarrow0^+0^-$ & $ D^*_{s0}(2317)^{+}K^{-}$ & $0.18\%$ & $0.46\%$ & $0.46\%$&$\square$&0.24\%\\
\rule[-2mm]{0mm}{6.5mm}
$1^+\rightarrow1^-1^-$ & $ D_s^{*+}K^{*}$ & $-$ & $-$ & $1.15\%$&$-$&$-$\\
\rule[-2mm]{0mm}{6.5mm}
$1^+\rightarrow1^+0^-$ & $ D_{s1}(2460)^{+}K^{-}$ & $0.02\%$ & $0.06\%$ & $0.004\%$&$\square$&0.00004\%\\
\rule[-2mm]{0mm}{6.5mm}
$$ & $ D_{s1}(2536)^{+}K^{-}$ & $-$ & $-$ & $0.01\%$&$-$&$-$\\
\hline
\rule[-2mm]{0mm}{6.5mm}
{\rm Total} & $$ & $1$ & $1$ & $1$&1&1\\
 \hline\hline
\end{longtable}
\end{center}
\end{table}

\begin{table}
\begin{center}
\linespread{0.5}
\begin{longtable}{ccccccc}
\caption{Branching ratios of different decay channels of $D_J(3000)^0$ as the 2P$(1^{+})$ state.} \\
\hline\hline
\rule[-2mm]{0mm}{6.5mm}{}
&{Final state}&{\bfseries Ours}&{\bfseries Ref.\cite{liuxiang3000}}&{\bfseries Ref.\cite{wangzhigang3000}}&{\bfseries Ref.\cite{Godfreyresults}}&{\bfseries Ref. \cite{lidemin3000}}\\
\hline
\rule[-2mm]{0mm}{6.5mm}
$1^+\rightarrow1^-0^-$ & $ D^*(2007)^0\pi^0$ &  $1.04\%$ & $\multirow{2}{*}{2.32\%}$ &  $4.04\%$& $\multirow{2}{*}{30.32\%}$&$5.64\%$\\
\rule[-2mm]{0mm}{6.5mm}
$$&$ D^*(2010)^+\pi^-$ & $1.95\% $ & $$ &  $8.06\%$&&$11.27\%$\\
\rule[-2mm]{0mm}{6.5mm}
$$&$ D^*(2007)^0\eta$ & $0.11\%$ & $0.87\%$&  $0.84\%$& $4.00\%$&$2.48\%$\\
\rule[-2mm]{0mm}{6.5mm}
$$&$ D^*(2007)^0\eta'$ & $0.08\%$ & $0.00046\%$& $6.38\%$& $\square$&$0.34\%$\\
\rule[-2mm]{0mm}{6.5mm}
$$ & $ D^*(2600)^0\pi^0$ & $5.10\%$ & $\square$ & $\square$ & $\multirow{2}{*}{1.04\%}$ &  $2.14\%$ \\
\rule[-2mm]{0mm}{6.5mm}
$$ & $ D^*(2600)^+\pi^-$ & $10.21\%$ & $\square$ & $\square$ & &$4.27\%$\\
\rule[-2mm]{0mm}{6.5mm}
$$ & $ D^*(2650)^0\pi^0$ & $0.28\%$ & $\square$ & $\square$ & $\square$ &$0.007\%$\\
\rule[-2mm]{0mm}{6.5mm}
$$ & $D^*(2650)^+\pi^-$& $0.55\%$ & $\square$ & $\square$ & $\square$ &$0.01\%$\\
\rule[-2mm]{0mm}{6.5mm}
$1^+\rightarrow0^-1^-$ & $ D^{0}\rho^{0}$ & $2.03\%$ & $\multirow{2}{*}{8.39\%}$& $5.89\%$&$\multirow{2}{*}{2.72\%}$&$0.42\%$\\
\rule[-2mm]{0mm}{6.5mm}
$$&$ D^{+}\rho^{-}$ & $4.60\%$ &$$& $11.77\%$&&$0.71\%$\\
\rule[-2mm]{0mm}{6.5mm}
$$&$ D^{0}\omega$ & $2.02\%$ & $2.67\%$ & $5.90\%$&0.88\%&$0.34\%$\\
\rule[-2mm]{0mm}{6.5mm}
$1^+\rightarrow1^-1^-$ & $ D^{*}(2007)^0\rho^{0}$ & $12.07\%$ & $\multirow{2}{*}{25.00\%}$ & $6.29\%$&$\multirow{2}{*}{19.52\%}$&$11.85\%$\\
\rule[-2mm]{0mm}{6.5mm}
$$&$ D^{*}(2010)^+\rho^{-}$ & $22.74\%$ &$$& $12.37\%$&&$22.61\%$\\
\rule[-2mm]{0mm}{6.5mm}
$$&$ D^{*}(2007)^0\omega$ & $11.64\%$ & $8.21\%$ & $5.98\%$& 6.56\%&$11.30\%$ \\
\rule[-2mm]{0mm}{6.5mm}
$1^+\rightarrow0^+0^-$ & $ D^*_0(2400)^{0}\pi^{0}$ & $0.49\%$ & $\multirow{2}{*}{19.64\%}$ & $\multirow{2}{*}{0.058\%}$ &$\multirow{2}{*}{3.92\%}$&$0.34\%$\\
\rule[-2mm]{0mm}{6.5mm}
$$&$ D^*_0(2400)^{+}\pi^{-}$ & $1.49\%$ & $$ & $$ &$$&$0.71\%$\\
\rule[-2mm]{0mm}{6.5mm}
$$&$ D^*_0(2400)^{0}\eta$ & $0.24\%$ & $0.25\%$ & $0.10\%$&$\square$&$0.14\%$\\
\rule[-2mm]{0mm}{6.5mm}
$1^+\rightarrow1^+0^-$ & $ D_1(2420)^0\pi^0$ & $2.50\%$ & $\multirow{2}{*}{15.71\%}$ & $\multirow{2}{*}{0.008\%}$&$\multirow{2}{*}{4.16\%}$&$4.08\%$\\
\rule[-2mm]{0mm}{6.5mm}  
$$ & $ D_1(2420)^+\pi^-$ & $5.02\%$ & $$ & $$&&$8.16\%$\\
\rule[-2mm]{0mm}{6.5mm}
$$&$ D_1(2420)\eta$ & $0.003\%$ & $0.00411\%$ & $0.002\%$&$\square$&$0.01\%$\\
\rule[-2mm]{0mm}{6.5mm}
$$&$ D_1(2430)^0\pi^0$ & $0.16\%$ &$\multirow{2}{*}{9.46\%}$ & $\multirow{2}{*}{0.002\%}$&$\multirow{2}{*}{2.00\%}$&$0.05\%$\\
\rule[-2mm]{0mm}{6.5mm}
$$&$ D_1(2430)^+\pi^-$ & $0.32\%$ &$$ & $$&&$0.10\%$\\
\rule[-2mm]{0mm}{6.5mm}
$$&$ D_1(2430)^0\eta$ & $-$ &$-$ & $0.001\%$&$-$&$-$\\
\rule[-2mm]{0mm}{6.5mm}
$1^+\rightarrow2^+0^-$ & $ D_2^*(2460)^{0}\pi^{0}$ & $3.08\%$ & $\multirow{2}{*}{5.89\%}$ & $9.57\%$&$\multirow{2}{*}{5.92\%}$&$2.52\%$\\
\rule[-2mm]{0mm}{6.5mm}
$$&$ D_2^*(2460)^{+}\pi^{-}$ & $6.06\%$ & $$ & $19.18\%$&&$4.94\%$\\
\rule[-2mm]{0mm}{6.5mm}
$$&$ D_2^*(2460)^{0}\eta$ & $-$ & $-$ & $0.19\%$&$-$&$-$\\
\hline
\rule[-2mm]{0mm}{6.5mm}
$1^+\rightarrow0^-1^-$ & $ D_s^{+}K^{*-}$ & $0.44\%$ & $1.25\%$ & $1.30\%$& 11.44\%&$3.76\%$ \\
\rule[-2mm]{0mm}{6.5mm}
$1^+\rightarrow1^-0^-$ & $ D_s^{*+}K^-$ & $0.06\%$ & $0.17\%$ & $0.41\%$&7.2\%&$1.49\%$ \\
\rule[-2mm]{0mm}{6.5mm}
$1^+\rightarrow0^+0^-$ & $ D^*_{s0}(2317)^{+}K^{-}$ & $5.65\%$ & $2.14\%$ & $0.17\%$&$\square$&$0.27\%$\\
\rule[-2mm]{0mm}{6.5mm}
$1^+\rightarrow1^-1^-$ & $ D_s^{*+}K^{*}$ & $-$ & $-$ & $1.39\%$&$-$&$-$\\
\rule[-2mm]{0mm}{6.5mm}
$1^+\rightarrow1^+0^-$ & $ D_{s1}(2460)^{+}K^{-}$ & $0.04\%$ & $0.08\%$ & $0.008\%$&$\square$&$0.004\%$\\
\rule[-2mm]{0mm}{6.5mm}
$$ & $ D_{s1}(2536)^{+}K^{-}$ & $-$ & $-$ & $0.016\%$&$-$&$-$\\
\hline
\rule[-2mm]{0mm}{6.5mm}
{\rm Total} & $$ & $1$ & $1$ & $1$&1&1\\
 \hline\hline
\end{longtable}
\end{center}
\end{table}

\subsection{For $D_{sJ}(3040)$}

For $D_{sJ}(3040)$, Tables.~VI and VII show the decay widths of $D_{sJ}(3040)^+$ as 2P$(1^{+'})$ and 2P$(1^{+})$ states, respectively. Also, the branching ratios are given in Tables.~VIII and IX. 

In our results, the first thing we notice is that the total widths are 157.4 MeV and 63.5 MeV for the 2P$(1^{+'})$ and 2P$(1^{+})$ states, and the former one is very close to the lower limit in experiments. Moreover, $D_{sJ}(3040)^+$ was discovered in the $D^*K$ channel. In our results, this channel has the largest branching ratio, taking up $60\%$ for the 2P$(1^{+'})$ state, but this channel only takes up less than $1\%$ if $D_{sJ}(3040)^+$ is the 2P$(1^{+})$ state. Therefore, our assumption of 2P$(1^{+'})$ state is more reasonable. In experiments, only the $D^*K$ channel has been observed yet, while there are many more channels for $D_{sJ}(3040)^+$. So we encourage more precise detection of $D^*K^*$ and $DK^*$. These two decay channels have the second and third largest branching ratios, respectively, and their ratio is $1:0.40$ in our calculation. 

In Ref. \cite{Godfreyresults}, the results in the G-I model were 147.6 MeV and 143.0 MeV for the 2P$(1^{+'})$ and 2P$(1^{+})$ states, respectively, and the largest channel was $D^*K$ for both assignments. The most noticeable difference was the $DK^*$, which accounted for $21.7\%$ and $4.6\%$, respectively in these two assumptions. In Ref. \cite{Lidemin3040}, the total widths were slightly larger than the upper limit and smaller than the lower limit for these two assignments, so the authors concluded that both two assignments seemed to be the quantum state for $D_{sJ}(3040)^+$.

In addition, there have been other studies involving the strong decay of $D_{sJ}(3040)^+$. In Ref. \cite{liuxiang3040}, Liu $et$ $al.$ employed the QPC model to calculate the partial and total decay widths of $D_{sJ}(3040)^+$ as the function of value $R$, which was chosen to reproduce the root mean square (rms) radius obtained by solving the Schr{\"o}dinger equation with the linear potential; they concluded that 2P$(1^{+'})$ was suitable. Liu $et$ $al.$ also used a modified G-I model to calculate these strong decays, and they gave the results of 285.83 MeV and 131.28 MeV for the 2P$(1^{+'})$ and 2P$(1^{+})$ states \cite{liuxiang3040modifiedGI}. In Ref. \cite{ZhongXianHui3040}, the authors drew the figure of decay widths as functions of the mixing angle. At the mixing angle $\phi \approx -54.7^{\circ}$ in the heavy quark limit, the total decay width is around 160 MeV for the 2P$(1^{+'})$ state, so the authors favored the 2P$(1^{+'})$ state. This result is consistent with ours. Moreover, in Ref. \cite{twodecaychannels3040}, the decay widths of $DK^*$ and $D_s\phi$ were given at around 95 MeV and 44 MeV, respectively, for the 2P$(1^{+'})$ state, which are larger than the corresponding results in any other model. In Ref. \cite{3040n3n4}, the total decay widths were 432.54 MeV and 301.52 MeV for the 2P$(1^{+'})$ and 2P$(1^{+})$ states, respectively, which are larger than the widths in other models.

We notice that since this resonance was first reported by BABAR in 2009 with a large error bar, there has not been any other update in experiments. So we call for more precise detection in experiments for the mass, total width, and strong decay properties.

\begin{center}
\begin{table*}[h]
\caption{Partial and total decay widths (in units of MeV) of $D_{sJ}(3040)^+$ as the 2P$(1^{+'})$ state.}
\begin{center}
\begin{tabular}{ccccccc}
\hline\hline
\rule[-2mm]{0mm}{6.5mm}{}&{Final state}&{\bfseries Ours}&{\bfseries Ref.\cite{Godfreyresults}}&{\bfseries Ref.\cite{Lidemin3040}}\\
\hline
\rule[-2mm]{0mm}{6.5mm}
$1^+\rightarrow1^-0^-$ & $ D^{*}(2007)^0K^+$ & $48.06$ &\multirow{2}{*}{36.5} & $34.35$ \\
\rule[-2mm]{0mm}{6.5mm}
$$&$ D^{*}(2010)^+K^0$ & $47.00$& & $34.84$\\
\rule[-2mm]{0mm}{6.5mm}
$1^+\rightarrow0^+0^-$ &$  D^{*}_0(2400)^0K^+$ & $3.71$ &\multirow{2}{*}{1.14}& $19.07$\\
\rule[-2mm]{0mm}{6.5mm}
$$&$ D^{*}_0(2400)^+K^0$ & $3.74$& & $14.39$\\
\rule[-2mm]{0mm}{6.5mm}
$1^+\rightarrow2^+0^-$ & $ D_2^*(2460)^0K^+$ & $2.83$ &\multirow{2}{*}{28.4} & $39.68$\\
\rule[-2mm]{0mm}{6.5mm}
$$&$ D_2^*(2460)^+K^0$ & $4.87$ & &$38.97$\\
\rule[-2mm]{0mm}{6.5mm}
$1^+\rightarrow1^-1^-$&$  D^*(2007)^0K^{*+}$ & $12.67$&\multirow{2}{*}{29.7} & $34.59$ \\
\rule[-2mm]{0mm}{6.5mm}
$$ & $ D^*(2010)^+K^{*0}$ & $11.77$& & $32.24$\\
\rule[-2mm]{0mm}{6.5mm}
$1^+\rightarrow0^-1^-$&$ D^0K^{*+}$ & $5.05$& \multirow{2}{*}{32.1} &$31.85$\\
\rule[-2mm]{0mm}{6.5mm}
$$&$ D^+K^{*0}$ & $4.78$ && $30.31$\\
\rule[-2mm]{0mm}{6.5mm}
$1^+\rightarrow1^+0^-$ & $ D_1(2420)^0K^+$ & $2.73$ &\multirow{2}{*}{12.2} & $1.76$\\
\rule[-2mm]{0mm}{6.5mm}
$$&$  D_1(2420)^+K^0$ & $2.67$& & $1.77$\\
\rule[-2mm]{0mm}{6.5mm}
$$&$ D_1(2430)^0K^+$ & $1.58$ &\multirow{2}{*}{3.38}&$0.5$ \\
\rule[-2mm]{0mm}{6.5mm}
$$&$ D_1(2430)^+K^0$ & $1.24$ &&$0.48$\\
\hline
\rule[-2mm]{0mm}{6.5mm}
$1^+\rightarrow1^-0^-$ & $ D_s^{*+}\eta$ & $4.22$ &0.153& $6.20$ \\
\rule[-2mm]{0mm}{6.5mm}
$1^+\rightarrow0^+0^-$&$ D^*_{s0}(2317)^{+}\eta$ & $0.37$&$\square$ & $3.12$\\
\rule[-2mm]{0mm}{6.5mm}
$1^+\rightarrow1^+0^-$&$ D_{s1}(2460)^{+}\eta$ & $0.07$&$\square$ & $0.03$ \\
\rule[-2mm]{0mm}{6.5mm}
$1^+\rightarrow1^+1^-$&$ D_{s}^{+}\phi$ & $\square$&4.15& $0.39$\\
\hline
\rule[-2mm]{0mm}{6.5mm}
${\rm Total}$ & $Exp:239\pm35^{+46}_{-42}$ & $157.4$ &147.6& $324.5$ \\
 \hline\hline
\end{tabular}
\end{center}
\end{table*}
\end{center}

\begin{center}
\begin{table*}[h]
\caption{Partial and total decay widths (in units of MeV) of $D_{sJ}(3040)^+$ as the 2P$(1^{+})$ state.}
\begin{center}
\begin{tabular}{ccccccc}
\hline\hline
\rule[-2mm]{0mm}{6.5mm}{}&{Final state}&{\bfseries Ours}&{\bfseries Ref.\cite{Godfreyresults}}&{\bfseries Ref.\cite{Lidemin3040}}\\
\hline
\rule[-2mm]{0mm}{6.5mm}
$1^+\rightarrow1^-0^-$ & $ D^{*}(2007)^0K^+$ & $0.02$ &\multirow{2}{*}{61.3} & $7.99$ &  $\multirow{2}{*}{}$&$$\\
\rule[-2mm]{0mm}{6.5mm}
$$&$ D^{*}(2010)^+K^0$ & $0.02$& & $7.79$ &  $$&$$\\
\rule[-2mm]{0mm}{6.5mm}
$1^+\rightarrow0^+0^-$ &$  D^{*}_0(2400)^0K^+$ & $3.46$ &\multirow{2}{*}{4.95}& $6.86$&  $\multirow{2}{*}{}$ & $$\\
\rule[-2mm]{0mm}{6.5mm}
$$&$ D^{*}_0(2400)^+K^0$ & $3.86$& & $6.43$&  $$& $$\\
\rule[-2mm]{0mm}{6.5mm}
$1^+\rightarrow2^+0^-$ & $ D_2^*(2460)^0K^+$ & $1.05$ &\multirow{2}{*}{0.67} & $3.00$& $\multirow{2}{*}{}$& $$\\
\rule[-2mm]{0mm}{6.5mm}
$$&$ D_2^*(2460)^+K^0$ & $1.96$ & &$2.89$& $$& $$\\
\rule[-2mm]{0mm}{6.5mm}
$1^+\rightarrow1^-1^-$&$  D^*(2007)^0K^{*+}$ & $17.06$&\multirow{2}{*}{38.9} & $39.84$ & $\multirow{2}{*}{}$&$$\\
\rule[-2mm]{0mm}{6.5mm}
$$ & $ D^*(2010)^+K^{*0}$ & $15.81$& & $37.36$ & $$&$$\\
\rule[-2mm]{0mm}{6.5mm}
$1^+\rightarrow0^-1^-$&$ D^0K^{*+}$ & $4.83$& \multirow{2}{*}{6.54} &$12.74$& $\multirow{2}{*}{}$&$$\\
\rule[-2mm]{0mm}{6.5mm}
$$&$ D^+K^{*0}$ & $4.47$ && $13.27$ & $$&$$\\
\rule[-2mm]{0mm}{6.5mm}
$1^+\rightarrow1^+0^-$ & $ D_1(2420)^0K^+$ & $2.75$ &\multirow{2}{*}{3.52} & $4.99$ & $\multirow{2}{*}{}$&$$\\
\rule[-2mm]{0mm}{6.5mm}
$$&$  D_1(2420)^+K^0$ & $2.7$& & $5.01$ & $$&$$\\
\rule[-2mm]{0mm}{6.5mm}
$$&$ D_1(2430)^0K^+$ & $0.08$ &\multirow{2}{*}{1.29}&$1.59$ & $\multirow{2}{*}{}$&$$\\
\rule[-2mm]{0mm}{6.5mm}
$$&$ D_1(2430)^+K^0$ & $0.05$ &&$1.52$ & $$&$$\\
\hline
\rule[-2mm]{0mm}{6.5mm}
$1^+\rightarrow1^-0^-$ & $ D_s^{*+}\eta$ & $3.77$ &9.65& $1.10$ & $$&$$\\
\rule[-2mm]{0mm}{6.5mm}
$1^+\rightarrow0^+0^-$&$ D^*_{s0}(2317)^{+}\eta$ & $1.56$&$\square$ & $1.19$ & $$&$$\\
\rule[-2mm]{0mm}{6.5mm}
$1^+\rightarrow1^+0^-$&$ D_{s1}(2460)^{+}\eta$ & $0.03$&$\square$ & $0.10$ & \\
\rule[-2mm]{0mm}{6.5mm}
$1^+\rightarrow1^+1^-$&$ D_{s}^{+}\phi$ & $\square$&16.2& $0.40$ \\
\hline
\rule[-2mm]{0mm}{6.5mm}
{\rm Total} & $Exp:239\pm35^{+46}_{-42}$ & $63.5$ &143.0& $154.1$ & $$ & $$ \\
 \hline\hline
\end{tabular}
\end{center}
\label{Decay widths of $D_{sJ}(3040)^+$ as 2P$(1^{+})$ state(MeV)}
\end{table*}
\end{center}

\begin{center}
\begin{table*}[h]
\caption{Branching ratios of different decay channels of $D_{sJ}(3040)^+$ as the 2P$(1^{+'})$ state.}
\begin{center}
\begin{tabular}{ccccccc}
\hline\hline
\rule[-2mm]{0mm}{6.5mm}{}&{Final state}&{\bfseries Ours}&{\bfseries Ref.\cite{Godfreyresults}}&{\bfseries Ref.\cite{Lidemin3040}}\\
\hline
\rule[-2mm]{0mm}{6.5mm}
$1^+\rightarrow1^-0^-$ & $ D^{*}(2007)^0K^+$ &  $30.54\%$ &$\multirow{2}{*}{24.7\%}$ & $10.58\%$\\
\rule[-2mm]{0mm}{6.5mm}
$$&$ D^{*}(2010)^+K^0$ & $29.86\%$ & & $10.73\%$\\
\rule[-2mm]{0mm}{6.5mm}
$1^+\rightarrow0^+0^-$ &$  D^{*}_0(2400)^0K^+$ & $2.35\%$ &$\multirow{2}{*}{0.772\%}$& $5.87\%$\\
\rule[-2mm]{0mm}{6.5mm}
$$&$ D^{*}_0(2400)^+K^0$ & $2.37\%$ && $4.43\%$\\
\rule[-2mm]{0mm}{6.5mm}
$1^+\rightarrow2^+0^-$ & $ D_2^*(2460)^0K^+$ & $1.79\%$ &$\multirow{2}{*}{19.2\%}$& $12.22\%$\\
\rule[-2mm]{0mm}{6.5mm}
$$&$ D_2^*(2460)^+K^0$ & $3.09\%$ &&$12.01\%$\\
\rule[-2mm]{0mm}{6.5mm}
$1^+\rightarrow1^-1^-$&$  D^*(2007)^0K^{*+}$ & $8.05\%$ &$\multirow{2}{*}{20.1\%}$& $10.65\%$\\
\rule[-2mm]{0mm}{6.5mm}
$$ & $ D^*(2010)^+K^{*0}$ & $7.47\%$ && $9.93\%$\\
\rule[-2mm]{0mm}{6.5mm}
$1^+\rightarrow0^-1^-$&$ D^0K^{*+}$ & $3.20\%$ &$\multirow{2}{*}{21.7\%}$&$9.81\%$\\
\rule[-2mm]{0mm}{6.5mm}
$$&$ D^+K^{*0}$ & $3.03\%$ && $9.33\%$\\
\rule[-2mm]{0mm}{6.5mm}
$1^+\rightarrow1^+0^-$ & $ D_1(2420)^0K^+$ & $1.73\%$ &$\multirow{2}{*}{8.26\%}$& $0.54\%$\\
\rule[-2mm]{0mm}{6.5mm}
$$&$  D_1(2420)^+K^0$ & $1.69\%$ && $5.54\%$\\
\rule[-2mm]{0mm}{6.5mm}
$$&$ D_1(2430)^0K^+$ & $1.00\%$ &$\multirow{2}{*}{2.29\%}$& $0.15\%$\\
\rule[-2mm]{0mm}{6.5mm}
$$&$ D_1(2430)^+K^0$ & $0.78\%$ && $0.15\%$\\
\hline
\rule[-2mm]{0mm}{6.5mm}
$1^+\rightarrow1^-0^-$ & $ D_s^{*+}\eta$ & $2.68\%$ &0.104\%& $1.91\%$\\
\rule[-2mm]{0mm}{6.5mm}
$1^+\rightarrow0^+0^-$&$ D^*_{s0}(2317)^{+}\eta$ & $0.23\%$ &$\square$& $0.96\%$\\
\rule[-2mm]{0mm}{6.5mm}
$1^+\rightarrow1^+0^-$&$ D_{s1}(2460)^{+}\eta$ & $0.04\%$ &$\square$& $0.009\%$\\
\rule[-2mm]{0mm}{6.5mm}
$1^+\rightarrow1^+1^-$&$ D_{s}^{+}\phi$ & $\square$ &2.81\%& $0.12\%$\\
\hline
\rule[-2mm]{0mm}{6.5mm}
Total & $$ & $1$ &$1$& $1$\\
 \hline\hline
\end{tabular}
\end{center}
\label{Branching ratios of $D_{sJ}(3040)^+$ as 2P$(1^{+'})$ state}
\end{table*}
\end{center}

\begin{center}
\begin{table*}[h]
\caption{Branching ratios of different decay channels of $D_{sJ}(3040)^+$ as the 2P$(1^{+})$ state.}
\begin{center}
\begin{tabular}{ccccccc}
\hline\hline
\rule[-2mm]{0mm}{6.5mm}{}&{Final state}&{\bfseries Ours}&{\bfseries Ref.\cite{Godfreyresults}}&{\bfseries Ref.\cite{Lidemin3040}}\\
\hline
\rule[-2mm]{0mm}{6.5mm}
$1^+\rightarrow1^-0^-$ & $ D^{*}(2007)^0K^+$ &  $0.03\%$ &$\multirow{2}{*}{42.87\%}$ & $5.19\%$\\
\rule[-2mm]{0mm}{6.5mm}
$$&$ D^{*}(2010)^+K^0$ & $0.03\%$ & & $5.06\%$\\
\rule[-2mm]{0mm}{6.5mm}
$1^+\rightarrow0^+0^-$ &$  D^{*}_0(2400)^0K^+$ & $5.45\%$ &$\multirow{2}{*}{3.46\%}$& $4.45\%$\\
\rule[-2mm]{0mm}{6.5mm}
$$&$ D^{*}_0(2400)^+K^0$ & $6.08\%$ && $4.17\%$\\
\rule[-2mm]{0mm}{6.5mm}
$1^+\rightarrow2^+0^-$ & $ D_2^*(2460)^0K^+$ & $1.65\%$ &$\multirow{2}{*}{0.47\%}$& $1.95\%$\\
\rule[-2mm]{0mm}{6.5mm}
$$&$ D_2^*(2460)^+K^0$ & $3.08\%$ &&$1.87\%$\\
\rule[-2mm]{0mm}{6.5mm}
$1^+\rightarrow1^-1^-$&$  D^*(2007)^0K^{*+}$ & $26.87\%$ &$\multirow{2}{*}{27.21\%}$& $25.86\%$\\
\rule[-2mm]{0mm}{6.5mm}
$$ & $ D^*(2010)^+K^{*0}$ & $24.91\%$ && $24.25\%$\\
\rule[-2mm]{0mm}{6.5mm}
$1^+\rightarrow0^-1^-$&$ D^0K^{*+}$ & $7.61\%$ &$\multirow{2}{*}{4.57\%}$&$8.27\%$\\
\rule[-2mm]{0mm}{6.5mm}
$$&$ D^+K^{*0}$ & $7.04\%$ && $8.61\%$\\
\rule[-2mm]{0mm}{6.5mm}
$1^+\rightarrow1^+0^-$ & $ D_1(2420)^0K^+$ & $4.33\%$ &$\multirow{2}{*}{2.46\%}$& $3.24\%$\\
\rule[-2mm]{0mm}{6.5mm}
$$&$  D_1(2420)^+K^0$ & $4.25\%$ && $3.25\%$\\
\rule[-2mm]{0mm}{6.5mm}
$$&$ D_1(2430)^0K^+$ & $0.13\%$ &$\multirow{2}{*}{0.90\%}$& $1.03\%$\\
\rule[-2mm]{0mm}{6.5mm}
$$&$ D_1(2430)^+K^0$ & $0.08\%$ && $0.99\%$\\
\hline
\rule[-2mm]{0mm}{6.5mm}
$1^+\rightarrow1^-0^-$ & $ D_s^{*+}\eta$ & $5.94\%$ &6.75\%& $0.71\%$\\
\rule[-2mm]{0mm}{6.5mm}
$1^+\rightarrow0^+0^-$&$ D^*_{s0}(2317)^{+}\eta$ & $2.46\%$ &$\square$& $0.77\%$\\
\rule[-2mm]{0mm}{6.5mm}
$1^+\rightarrow1^+0^-$&$ D_{s1}(2460)^{+}\eta$ & $0.05\%$ &$\square$& $0.06\%$\\
\rule[-2mm]{0mm}{6.5mm}
$1^+\rightarrow1^+1^-$&$ D_{s}^{+}\phi$ & $\square$ &11.33\%& $0.26\%$\\
\hline
\rule[-2mm]{0mm}{6.5mm}
Total & $$ & $1$ &$1$& $1$\\
 \hline\hline
\end{tabular}
\end{center}
\label{Branching ratios of $D_{sJ}(3040)^+$ as 2P$(1^{+})$ state}
\end{table*}
\end{center}

\section{SUMMARY}

The strong decay properties of $D_{J}(3000)$ and $D_{sJ}(3040)$ have been studied in this work. We have employed our instantaneous Bethe-Salpeter method to give the wave function of heavy-light mesons. Our calculation show that $D_{J}(3000)$ is a good candidate for the 2P$(1^{+'})$ state. Apart from $D^*\pi$, the $D_2^*(2460)\pi$ and $D^*(2600)\pi$ channels also have large partial decay widths, which are helpful in investigating the properties of $D_J(3000)$. For $D_{sJ}(3040)$, although our result is smaller than the central value and very close to the lower limit of the experimental data, we still treat it as a potential candidate for the 2P$(1^{+'})$ state, considering results in other assignments deviate from experimental data much more. Due to the large uncertainty in experiments and great differences between the predictions of different models, we call for more precise detections. Model-independent calculations, such as lattice QCD, can also provide a better and more comprehensive understanding of these newly discovered resonances.


 \section*{Acknowledgements}
This work was supported in part by the National Natural Science Foundation of China (NSFC)
under Grants No. 11505039, No. 11575048, No. 11405037, No. 11447601, No. 11535002, and No. 11675239, and in part by Program for Innovation Research of Science in Harbin Institute of Technology (PIRS of HIT) No. B201506 and No. A201409.

\bibliography{reference}


\end{document}